\documentclass[12pt]{article}
\usepackage{epsfig}
\usepackage{amssymb,amsmath}
\usepackage{setspace}

\usepackage{amsmath}

\newcommand{\bea}{\begin{eqnarray}}
\newcommand{\eea}{\end{eqnarray}}

\newcommand{\beq}{\begin{equation}}
\newcommand{\eeq}{\end{equation}}

 \makeatletter
\def\alt{\mathrel{\mathpalette\gl@align<}}
\def\agt{\mathrel{\mathpalette\gl@align>}}
\def\gl@align#1#2{\lower.6ex\vbox{\baselineskip\z@skip\lineskip\z@
\ialign{$\m@th#1\hfil##\hfil$\crcr#2\crcr\sim\crcr}}} \makeatother

\usepackage{longtable}

\begin{document}
%
\vspace*{1.0cm}

\begin{center}
\baselineskip 20pt {\Large\bf
 Inverse Seesaw  in NMSSM and 126 GeV Higgs Boson
}
\vspace{1cm}

{\large
Ilia Gogoladze\footnote{E-mail:ilia@bartol.udel.edu},
Bin He\footnote{ E-mail:hebin@udel.edu},
and Qaisar Shafi\footnote{ E-mail:shafi@bartol.udel.edu}
} \vspace{.5cm}

{\baselineskip 20pt \it
Bartol Research Institute, Department of Physics and Astronomy, \\
University of Delaware, Newark, DE 19716, USA \\
}
\vspace{.5cm}

\vspace{1.5cm} {\bf Abstract}
\end{center}

We consider extensions of the next-to-minimal supersymmetric model (NMSSM) in which the observed neutrino masses are
generated through a TeV scale inverse seesaw mechanism. The new particles associated with this mechanism can have sizable couplings to the Higgs field which can yield a large contribution to the mass of the lightest CP-even Higgs boson. With this new contribution, a 126 GeV Higgs is possible along with order of  200 GeV masses for the stop quarks for a broad range of $\tan\beta$.  The Higgs production and decay in the diphoton channel can be enhanced due to this new contribution. It is also possible to solve the little hierarchy problem in this model without invoking a maximal value for the NMSSM trilinear coupling and without  severe restrictions on the value of $\tan\beta$.

\thispagestyle{empty}

\newpage

\addtocounter{page}{-1}

\baselineskip 18pt

\section{Introduction}

The ATLAS and CMS Collaborations at the Large Hadron Collider (LHC)  have independently reported the discovery \cite{ATLAS:jul4,CMS:jul4} of a particle with production and decay modes that appear more or less consistent with the Standard Model (SM) Higgs boson of mass $m_h\approx126 $ GeV.
In addition to the Higgs discovery, both experiments reported an excess in Higgs production and decay in the diphoton channel, around $1.4-2$ times larger than the SM expectations. The statistical significance of this apparent deviation from the SM prediction is at present not sufficiently strong to draw a definite conclusion, but if confirmed in the future, it will be clear indication of new physics around the electroweak scale. These results nevertheless serve as strong motivation to investigate possible extensions of the SM where a possible signal in the diphoton channel could be enhanced compared to the SM.

The minimal supersymmetric standard model (MSSM) \cite{mssm} can accommodate values of $m_h \sim 126 \rm \ GeV$, but this  requires either a very large, ${\cal O} (\mathrm{few}-10)$ TeV,  stop quark mass \cite{Ajaib:2012vc}, or a large soft supersymmetry  breaking (SSB) trilinear A-term, with a stop quark mass of around a TeV \cite{Djouadi:2005gj}. Such a heavy stop quark
leads to the so-called ``little hierarchy" problem \cite{b5} because, in implementing radiative electroweak
symmetry breaking, TeV scale quantities must conspire to yield the electroweak mass scale.

On the other hand,  in the next-to-minimal supersymmetric standard model (NMSSM), the Higgs mass can be raised significantly through a tree level contribution to the Higgs potential \cite{Ellwanger:2009dp}. Therefore, the NMSSM can alleviate the little hierarchy problem, and a 126 GeV Higgs mass can be realized with less fine-tuning.
In Ref.~\cite{natural} it was shown that in order to accommodate  a 126 GeV Higgs mass with only a few percent  fine-tuning, the NMSSM is pushed to the edge of its parameter space, with $\tan\beta\lesssim 2$ and $\lambda \sim 0.7$. Here $\tan\beta$ is the ratio of the vacuum expectation values (VEVs) of the up ($H_u$) and down ($H_d$) MSSM Higgs doublets. The  parameter $\lambda$ is the dimensionless coupling associated with the interaction $H_u H_d S$, where $S$ is a MSSM gauge singlet field.
Note that assuming non-universal gaugino masses at the GUT scale, one can also alleviate the little hierarchy problem \cite{Abe:2007kf}, but we will not discuss this possibility in this Letter.

Furthermore, in the framework of the NMSSM, Higgs production and decay in the diphoton  channel can be enhanced with respect to the SM prediction  due to the doublet-singlet mixing in the Higgs sector \cite{natural,Gunion:2012zd}.
It has been shown that to comply with the ATLAS and CMS results, a large stop mass still cannot be avoided. Besides, the  couplings ($\lambda$, $\kappa$, $y_t$) are all of $\mathcal{O}$(1) at the GUT scale, which are close to the Landau pole.\footnote{The possible impact of non-perturbative couplings has also received attention. For an example, see Ref.~\cite{nonperturb}.} Here $\kappa$ is the dimensionless coupling corresponding  to the  $S^3$ interaction and $y_t$ is the top Yukawa coupling.

Inspired by recent studies on the  NMSSM and the results from ATLAS and CMS, we consider an extension of the NMSSM
which has previously been used to explain the origin of neutrino masses.
In Ref. \cite{Gogoladze:2008wz}, in particular, it was shown that in the NMSSM  the observed neutrino masses and mixings
can be  described in terms of dimension six, rather than dimension five, operators. All such operators respect the discrete $Z_3$ symmetries of the model. The new particles associated with the inverse  seesaw mechanism \cite{InvSeesaw} can have sizable couplings to the Higgs boson, even with the seesaw scale of around a TeV. This, as we will show, enables the Higgs boson mass to be 126 GeV, without invoking sizable contributions from the stop quark as well as keeping the $\lambda$ and  $\kappa$ couplings relatively small. With relatively light stop quarks in the spectrum  one can enhance the diphoton production relative to the SM prediction \cite{Djouadi:1998az, Desai:2012qy, Ajaib:2012eb}.

The layout of this Letter is as follows. In Section \ref{model}, we briefly summarize  the NMSSM and the upper bound on the lightest CP-even Higgs boson mass. In
Section \ref{nmssm+singlet} we  present the NMSSM with inverse seesaw mechanism for the neutrinos.  In this section  a SM gauge singlet field is introduced to generate dimension six (and seven) operators for the neutrinos. We discuss how the
inverse seesaw mechanism affects the upper bound on the
lightest CP-even Higgs boson mass. In Section \ref{nmssm+triplet}, the inverse seesaw mechanism is generated through an $SU(2)_L$ triplet field and its impact on the Higgs mass is also considered. Our conclusions are summarized in Section \ref{conclusions}.

\section{Higgs Boson Mass in MSSM and  NMSSM\label{model}}


The NMSSM is obtained by adding to the  MSSM a gauge singlet chiral superfield $S$ (with even
$Z_2$  matter parity) and including the following superpotential terms:
\bea
 W \supset \lambda S H_u H_d +\frac{\kappa}{3} S^3,
\eea
where $\lambda$ and $\kappa$ are dimensionless constants, and $H_u$,
$H_d$ denote the MSSM Higgs doublets. A discrete $Z_3$ symmetry
under which $S$ carries a unit charge $\omega=e^{i 2 \pi/3}$ is
introduced in order to eliminate terms from the superpotential  that are linear and
quadratic in $S$, as well as the MSSM $\mu$ term.  We also need  the $Z_3$ symmetry to  forbid
dangerous tadpole terms in the potential which can revive the gauge hierarchy problem in the theory.
On the other hand, once the $S$  field develops  a VEV, the $Z_3$ symmetry is spontaneously broken which can cause the domain
wall problem. In order to circumvent this problem, as pointed out in Ref. \cite{Panagiotakopoulos:1998yw},
suitable higher dimensional
operators can be introduced in the superpotential  which explicitly break the $Z_3$ symmetry,
thereby lifting the degeneracy between three discrete
vacua. Note that these $Z_3$ violating higher dimensional operators
($S^7/M^4_{Pl}$), where $M_{Pl}$ denotes the Planck mass, are quite different in form from the effective seesaw operators
which we will  discuss  in this Letter. The higher dimensional operators which generate neutrino masses are $Z_3$ invariant.

In order to assign the $Z_3$ charges we require the presence of Yukawa couplings at the renormalizable level.
There are several possible $Z_3$ charge assignments for the matter superfields presented in Ref. \cite{Gogoladze:2008wz} that are consistent
with this requirement.  We consider those  cases (see Table \ref{table1}) which lead to the dimension six (inverse) seesaw operator for neutrinos. Later, we will briefly discuss the dimension seven seesaw operators and their implications for the Higgs boson mass. We employ the standard notation for the superfields in Table \ref{table1}. Family indices are omitted for simplicity.
\begin{table}[h]
\begin{center}
\setlength{\tabcolsep}{5pt}
\renewcommand{\arraystretch}{1.2}
\begin{tabular}{| c|  c| c|  c | c | c | c | c | c | c| }
\hline
& & $Q$ & $U^c$& $D^c$ & $L$ & $E^c$& $H_u$ &$H_d$ & $S$    \\
 \hline %
 $case \,I$ & $Z_3$ & 1 & $\omega^2$ & $\omega^2$ &  1& $\omega^2$ & $\omega$& $\omega$& $\omega$\\[0.1cm]
 \hline
 $case \, II$ & $Z_3$ &  1 & $\omega$ & $1$ & $\omega^2$ & $\omega$ & $\omega^2$& $1$& $\omega$\\[0.1cm]
 \hline
 $case \, III$ & $Z_3$ &  1 & $1$ & $\omega$ &  $\omega$& $1$ & $1$& $\omega^2$& $\omega$\\[0.1cm]
 \hline
\end{tabular}
\end{center}
\caption{ $Z_3$ charge assignments of the NMSSM  superfields
corresponding to dimension  six
 operators for neutrino masses. Here $\omega=e^{i2\pi/3}$.
}
\label{table1}
\end{table}

The $Z_3$ charge assignments presented in Table \ref{table1} lead to the following effective operator for neutrino masses and mixing:
\begin{eqnarray}
 \frac{L L H_u H_u S}{M^2_6},
 \label{sw2}
\end{eqnarray}
where $M_{6}$ denotes the appropriate seesaw mass scale. As we will show in the next section, this operator  can be generated from the renormalizable superpotential by just integrating out the heavy (${\cal O}$(TeV)) fields. In section \ref{nmssm+singlet} we consider the gauge singlet case, and in section \ref{nmssm+triplet} we replace the gauge singlet field with an $SU(2)_L$ triplet field. We will also show later that the new TeV scale fields will affect the lightest  CP-even  Higgs mass bound. Before studying this new contribution to the lightest CP-even  Higgs boson mass, we briefly summarize the Higgs   mass bound in the MSSM and NMSSM.

The upper limit on the lightest CP-even Higgs boson mass in the NMSSM is given by \cite{nmssm}
\begin{eqnarray}
\left[ m_{h}^{2}\right] _{NMSSM} &=&M_{Z}^{2}\left( \cos ^{2}2\beta +\frac{2\lambda ^2}{g_1^2+g_2^2}\sin ^{2}2\beta \right)
\left( 1-\frac{3%
}{8\pi ^{2}} y_{t}^{2}t\right)  \nonumber \\
&+&\frac{3}{4\pi ^{2}}y_{t}^{2}m_{t}^{2}\sin ^{2}\beta \left[ \frac{1}{2}\widetilde{X}_{t}+t+\frac{%
1}{\left( 4\pi \right) ^{2}}\left(
\frac{3}{2}y_{t}^{2}-32\pi \alpha _{s}\right)
\left( \widetilde{X}_{t}+t\right)t \ \right],
\label{nmssm}
\end{eqnarray}
where
\begin{eqnarray}
t =\log \left(
\frac{M_{S}^{2}}{M_{t}^{2}}\right),~
\widetilde{X}_{t} &=&\frac{2\widetilde{A}_{t}^{2}}{M_{S}^{2}}\left( 1-\frac{\widetilde{A}%
_{t}^{2}}{12M_{S}^{2}}\right),~
\widetilde{A}_{t}=A_{t}-\lambda \langle S\rangle \cot \beta.
\label{A1}
\end{eqnarray}
$A_t$ is the top trilinear soft term, and  $\langle S\rangle$ denotes the vacuum expectation value (VEV) of the singlet field. Also, $g_1$ and $g_2$ denote the $U(1)_Y$ and the $SU(2)_L$ gauge couplings, $M_t=173.2$ GeV is the top quark pole mass, $M_S=\sqrt{m_{\tilde{t}_L}m_{\tilde{t}_R}}$ denotes the SUSY scale, and ${\tilde{t}_L}$ and  ${\tilde{t}_R}$ are the left and right handed stop quarks. Notice that we assume $\tan\beta< 50$, since for larger $\tan\beta$  values there can be additional contributions in Eq. (\ref{nmssm}) which may reduce the Higgs mass \cite{Ellwanger:2009dp}.
 An approximate error of $\pm 3$ GeV in the Higgs mass calculation is assumed, which largely arises from theoretical uncertainties \cite{Degrassi:2002fi} and simplifications in the calculation of the Higgs mass formula in Eq. (\ref{nmssm}).
 The upper bound on $\lambda$ at the weak scale depends on $\tan \beta$.  In general, it cannot be greater than $\sim 0.7$, if we require that $\lambda$ remains perturbative up to the $M_{\rm GUT}$ scale \cite{nmssm}.

Note that the main difference in the expression (see Eq. (\ref{nmssm})) for the  lightest CP-even Higgs mass between the NMSSM and the MSSM theory is the term ${2\lambda ^2}\sin ^{2}(2\beta)/(g_1^2+g_2^2)$. Therefore, the maximum value of the lightest CP-even Higgs mass in the NMSSM is obtained for smaller value of $\tan \beta$.

\begin{figure}[t]
\begin{center}
{\includegraphics*[width=.49\linewidth]{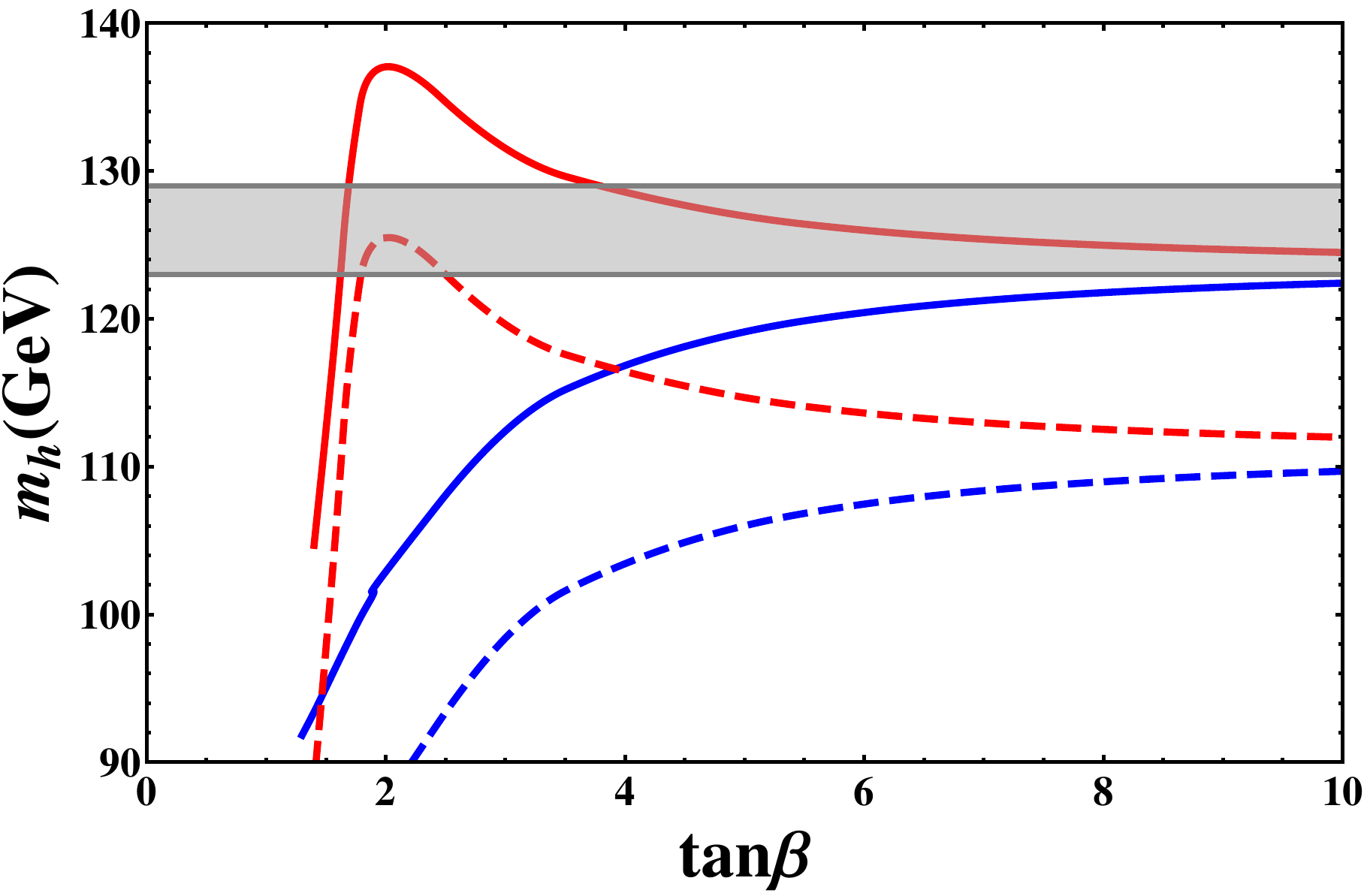}
\includegraphics*[width=.49\linewidth]{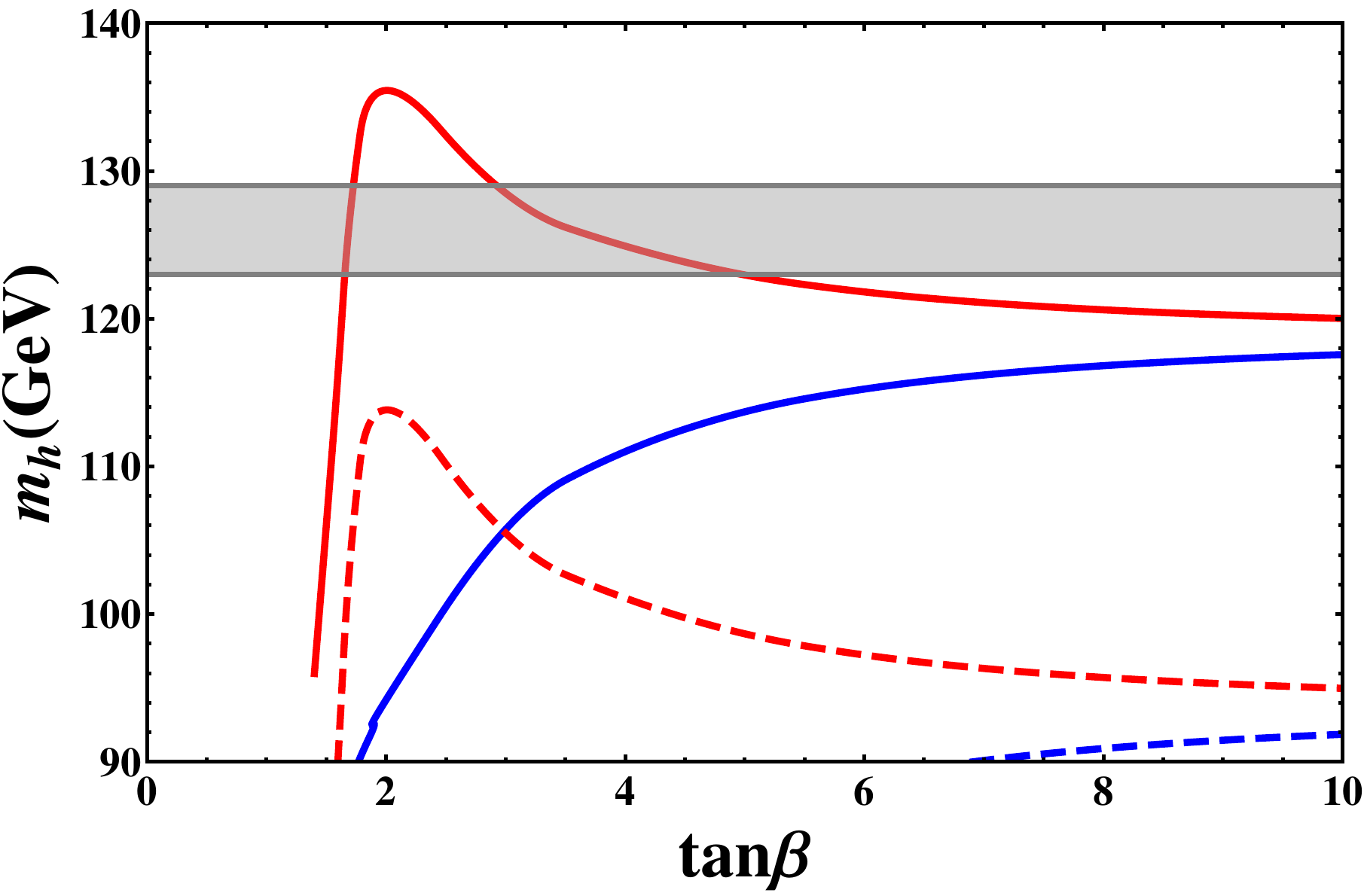}}
\caption{
Upper bounds on the lightest CP-even Higgs boson mass
versus $\tan \beta$, for $M_S=1$ TeV (left panel) and $M_S=200$ GeV (right panel). Maximum value of $\lambda$ is used.  Red lines correspond to the NMSSM, and blue lines correspond to the MSSM. The solid lines show the Higgs mass bounds for $\widetilde{X}_{t}=6$, while the dashed lines show the bounds with $\widetilde{X}_{t}=0$. The gray band shows the Higgs mass range of $126\pm 3$ GeV.
}
\label{fig1}
\end{center}
\end{figure}

In Figure \ref{fig1} we show our results in the $m_h$ versus $\tan\beta$ planes. For comparison, we have chosen two different SUSY scales,  $M_S=1$ TeV (left panel) and $M_S=200$ GeV (right panel), and the maximum value of the coupling $\lambda$ is used. The red lines correspond to the NMSSM case, whereas the blue lines correspond to the MSSM case. The solid lines show the Higgs mass bounds for $\widetilde{X}_{t}=6$, while the dashed lines show the bounds for $\widetilde{X}_{t}=0$. The gray band shows the Higgs mass range of $126\pm 3$ GeV. We can see that in order to obtain a 126 GeV Higgs in the MSSM, we need to have  $M_S>1$ TeV with maximal mixing.   In the NMSSM, due to the additional contributions proportional to  $\lambda$, for $\tan\beta=2$ one can easily get a 126 GeV Higgs mass even for $M_S=200$ GeV. However, without maximal mixing in the stop sector it is hard, even in the NMSSM, to generate a 126 GeV Higgs mass with $M_S< 1$ TeV.

\section{Inverse Seesaw and Higgs Boson Mass \label{seesaw}}
\subsection{NMSSM + Gauge Singlet field  \label{nmssm+singlet}}
As shown in Ref. \cite{Gogoladze:2008wz}, one can incorporate the observed solar and atmospheric neutrino oscillations in the NMSSM by introducing an effective dimension six operator for neutrino masses and mixings.
 The simplest way to generate this operator is to introduce the gauge singlet chiral superfields $(N_n^c+N_n)$ in the NMSSM with charges listed in Table \ref{tt2}. This charge assignment corresponds to the so-called  $case \,I$ in Table \ref{table1}.  It is straightforward to find the $Z_3$ charge assignments for $N_n^c+N_n$ for other cases given in Table \ref{table1}, but this will not lead to any new phenomena compared to $case \, I$. Because of this we will not consider here the other cases presented in Table \ref{table1}.  Since the new chiral superfields are gauge singlets, they will preserve gauge coupling unification which is one of the nice features of supersymmetry.
\begin{table}[h]
\begin{center}
\renewcommand{\arraystretch}{1.2}
\begin{tabular}{| c|  c|  c | c | c | c | c | c | c| }
\hline
   & $SU(3)_c$ & $SU(2)_L$   & $U(1)_Y$ & $Z_3$ & $Z_2$ \\
\hline
 $N^c_n$      &1  & 1  &   0  &  $\omega^2$  & $-$ \\
 \hline
 $N_n$         & 1 & 1  &   0 &  $\omega $    & $-$ \\
  \hline
\end{tabular}
\end{center}
\caption{Charge assignments of $N^c_n+N_n$  superfields for $case \,I$. Here  $\omega=e^{i2\pi/3}$, $n$ denotes  the number of  gauge singlet $(N_n^c+N_n)$   pairs, and $Z_2$ is matter parity.
}
\label{tt2}
\end{table}

The renormalizable superpotential terms involving only the new
chiral superfields are given by
\begin{eqnarray}
 W \supset  y_{ni}^N N_n^c (H_u L_i) + \frac{\lambda_{N_{nm}}}{2} S N_n N_m
     + m_{nm} N^c_n N_m.
     \label{eq6}
\end{eqnarray}
Here $i$ runs  from 1 to 3 and denotes the family index, while $n$ and $m$ denote the number of pairs of new fields which we consider, and can be from zero (just the NMSSM case)  up to 3. For $m_{nm}$  larger than the electroweak scale, we can  integrate out the $N^c_n$ and $N_n$ fields and generate the effective non-renormalizable  operators for neutrino masses presented in Eq. (\ref{sw2}).
Following the electroweak symmetry breaking,
 the neutrino Majorana mass matrix is generated:
\begin{eqnarray}
  m_\nu = \frac{(Y_N^T Y_N) {\it v_u}^2}{M_6} \times \frac{\lambda_N \langle S \rangle}{M_6}.
  \label{sw33}
\end{eqnarray}
For simplicity, we take $m_{ij} = M_6 \delta_{ij}$,
 $Y_{N} \equiv y_{ij}$, and $(\lambda_N)_{ij}= \lambda_N \delta_{ij}$.   ${\it v_u}$ is the VEV of the $H_u$ Higgs doublet and $\langle S \rangle$, the VEV of $S$ field, is around the TeV scale.\footnote{The smallness of $M_6$ can be understood using dimension 5 operator for mass generation. For an example, see Ref.~\cite{tjli}.} 
Eq. (\ref{sw33}) implies that even if $Y_N\sim{\cal O}(1)$ and
 $M_S \sim 1$ TeV, the correct mass scale for the  light neutrinos
 can be reproduced by suitably adjusting  $\lambda_N$.

From  Eq. (\ref{eq6}), the additional contribution to the lightest CP-even Higgs mass is given by
\begin{eqnarray}
\left[ m_{h}^{2}\right] _{N} &=&n \times \left[ -M_{Z}^{2}\cos ^{2}2\beta \left(
\frac{1}{8\pi ^{2}} Y_{N} ^{2}t_{N}\right)
+\frac{1}{4\pi ^{2}} Y_{N}^{4}v^{2}\sin ^{2}\beta \left( \frac{1
}{2}\widetilde{X}_{Y_{N}}+t_{N}\right) \right],
\label{e3}
\end{eqnarray}
where
\begin{eqnarray}
t_{N}=\log \left( \frac{%
M_{S}^{2}+M_{6}^{2}}{M_{6}^{2}}\right),~
\widetilde{X}_{Y_{N}}=\frac{4\widetilde{A}_{Y_{N}}^{2}\left(
3M_{S}^{2}+2M_{6}^{2}\right) -\widetilde{A}_{Y_{N}
}^{4}-8M_{S}^{2}M_{6}^{2}-10M_{S}^{4}}{6\left(
M_{S}^{2}+M_{6}^{2}\right) ^{2}},
\end{eqnarray}
and
\begin{eqnarray}
\widetilde{A}_{Y_{N}}=A_{Y_{N}}-Y_N \langle S\rangle \cot \beta.
\end{eqnarray}
$A_{Y_{N}}$ is the trilinear $N^{c}-L$ soft mixing parameter and $n$ is the number of pairs  of new singlets. $v=174.1$ GeV is the electroweak VEV. Note that the expression  in Eq. (\ref{eq6}) is very similar to what was presented in Ref. \cite{Babu:2004xg}.

\begin{figure}[t!]
\begin{center}
{\includegraphics*[width=.7\linewidth]{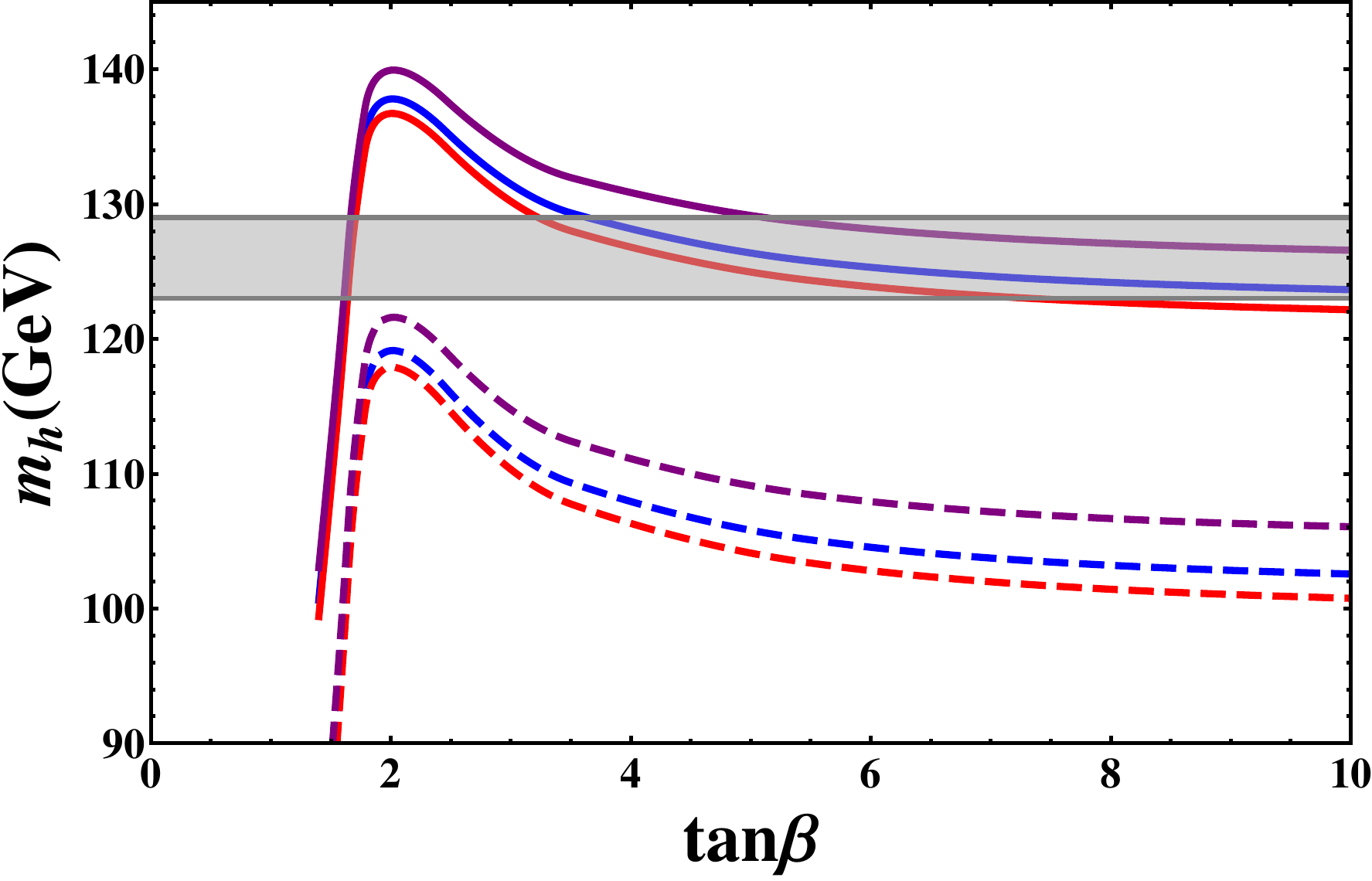}}
\caption{
Upper bounds on the lightest CP-even Higgs boson mass
versus $\tan \beta$, with $M_S=300$ GeV, $M_6=3$ TeV, $\widetilde{X}_{Y_N}=4$. Maximum value of $\lambda$ is used. Red lines correspond to NMSSM, while blue lines correspond to NMSSM with one additional pair of $(N_n^c+N_n)$ singlets. Purple  lines correspond to NMSSM with 3 additional pairs of $(N_n^c+N_n)$ singlets. In both cases $Y_N=0.7$. The solid lines show the Higgs mass bounds with $\widetilde{X}_{t}=6$, while the dashed lines show the bounds with $\widetilde{X}_{t}=0$. For reference the gray band shows the Higgs mass range of $126\pm 3$ GeV.
}
\label{fig2}
\end{center}
\end{figure}

To see how these new, $(N_n^c+N_n)$, singlets can affect the lightest CP-even Higgs mass, we plot the upper bounds on the lightest CP-even Higgs mass versus $\tan\beta$ for $n=1$ and 3 in Figure \ref{fig2}. We choose $M_S=300$ GeV for  all cases in order to minimize the stop quark contribution to the Higgs boson mass. $M_6=3$ TeV and $Y_N=0.7$ are used. Compared to the NMSSM bound, the Higgs mass can be increased by up to 5 GeV or so. To maximize the effect coming from the new field we choose the maximal value $\widetilde{X}_{Y_{N}}=4$.  For $n=3$, the upper bound for the Higgs mass becomes as large as 140 GeV for $\tan\beta\approx 2$, and asymptotically approaches  $m_h \approx 126$ GeV for large $\tan\beta$. This indicates that we are able to accommodate a Higgs mass of around 126 GeV even with relatively small values of $\lambda$ and $Y_N$. Therefore, we can conclude that in the NMSSM with the inverse seesaw mechanism for neutrinos, we can have relatively light ${\cal O}$(300) GeV  stop quarks. This can be achieved  without invoking maximal values for the  $\lambda$ or $Y_N$ couplings, and without imposing severe restrictions on the values of $\tan\beta$.

\begin{figure}[t!]
\begin{center}
{\includegraphics*[width=.7\linewidth]{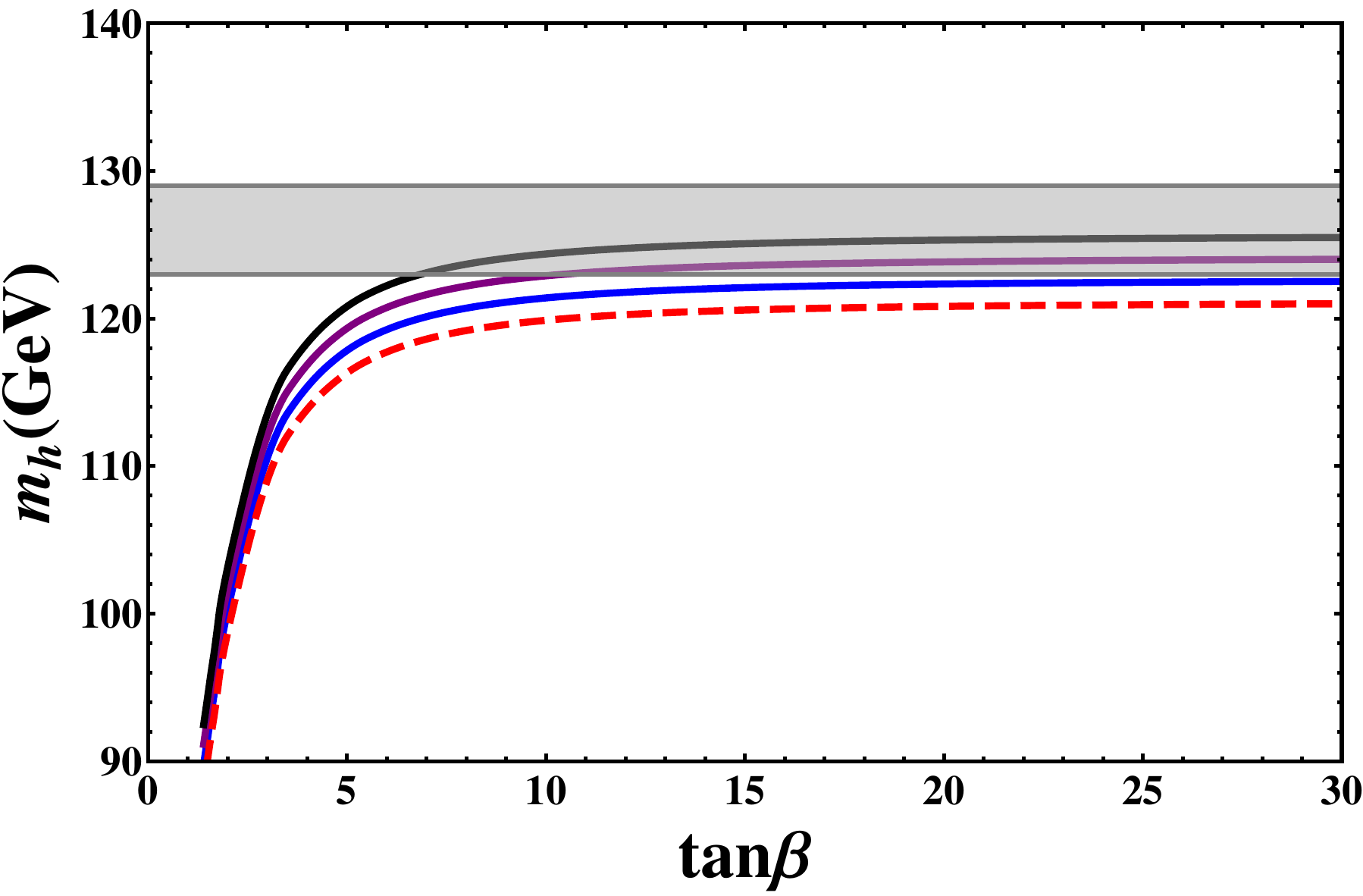}}
\caption{
Upper bounds on the lightest CP-even Higgs boson mass
versus $\tan \beta$, with $M_S=300$ GeV, $M_6=3$ TeV, $\widetilde{X}_{t}=6$, $\widetilde{X}_{Y_N}=4$, $Y_N=0.7$ and $\lambda=0.1$. Red dashed line corresponds to NMSSM. Blue, purple and black solid lines (from bottom to top) correspond to NMSSM+singlets with $n$=1, 2 and 3. For reference the gray band shows the Higgs mass range of $126\pm 3$ GeV.}
\label{fig3}
\end{center}
\end{figure}
In order to show how small the coupling $\lambda$ can be,  we consider the case with $\lambda=0.1$, $M_S=300$ GeV and $Y_N=0.7$. The main reason for choosing $\lambda=0.1$ is that in this case the contribution from $\lambda$ to the lightest CP-even Higgs mass is negligible, and the results are applicable to the MSSM case as well.    Figure \ref{fig3} shows the upper bounds on the Higgs mass versus $\tan\beta$ for varying numbers of $N_n^c+N_n$ pairs. Note that in order to reproduce the neutrino oscillation data, we  need to introduce at least two pairs of $N_n^c$ and $N_n$. However, for completeness, we have shown the bounds with $n=$1, 2 and 3 in Figure \ref{fig3}. We can see from Figure \ref{fig3} that in the MSSM and NMSSM, an inverse seesaw can make it very easy to generate $m_h=126$ GeV.
In this case we do not require very heavy stop quarks, or large value of $\lambda$, or a very restrictive value in the NMSSM of $\tan\beta \approx 2$ .

Table \ref{t3} presents upper bounds on the  Higgs masses  for varying numbers of ($N_n^c+N_n$) singlets. $n=0$ corresponds to NMSSM/MSSM without inverse seesaw.    The Higgs mass has  been calculated using the input values  $\tan \beta=30$,  $\lambda=0.1$, 
$\widetilde{X}_{t}=6$, $\widetilde{X}_{Y_N}=4$,
$Y_N=0.7$ and $M_S=300$ GeV.

\begin{table}[h]
\begin{center}
\renewcommand{\arraystretch}{1.2}
\begin{tabular}[h]{|c|c|c|c|c|c|}
\hline
 & $n=0$ & $n=1$&$n=2$ & $n=3$ \\
  \hline
 $m_h$(GeV) & 121 & 123& 124& 126  \\
  \hline
\end{tabular}
\end{center}
\caption{Higgs masses for varying numbers of $(N_n^c+N_n)$ singlets, with $n=0$ corresponding to NMSSM/MSMM. The Higgs mass has been calculated using the input values $\tan\beta=30$,  $\lambda=0.1$, $Y_N=0.7$, $\widetilde{X}_{t}=6$, $\widetilde{X}_{Y_N}=4$  and $M_S=300$ GeV.}
\label{t3}
\end{table}

As mentioned above, in order to have realistic neutrino masses and mixings, with TeV scale effective dimension six operators, we need to adjust the values for $\lambda_N$ in Eq. (\ref{sw33}). It turns out that $\lambda_N$
should be order of $10^{-9}$ or so, which is possible but appears not natural. This can be resolved if we consider $Z_3$  charge assignment which allows dimension seven as the lowest possible operator for generating neutrino masses. One example of such a charge assignment is presented in Table \ref{tt22}.

\begin{table}[h]
\begin{center}
\setlength{\tabcolsep}{5pt}
\renewcommand{\arraystretch}{1.2}
\begin{tabular}{| c|  c| c|  c | c | c | c | c | c | c|c|c| }
\hline
 & $Q$ & $U^c$& $D^c$ & $L$ & $E^c$& $H_u$ &$H_d$ & $S$ & $N^c_n$ & $N_n$ & $N_m^0$  \\
 \hline %
  $Z_3$ & 1 & $\omega$ & $\omega$ &   $\omega$ & $\omega$ & $\omega$& $\omega$& $\omega$ & $\omega$ & $\omega^2$& 1 \\[0.1cm]
 \hline
\end{tabular}
\end{center}
\caption{ $Z_3$ charge assignments of the NMSSM with additional new  superfields
 which correspond to  dimension  seven as  lowest effective
 operator for neutrino masses. The new fields have $Z_2$ matter parity and $\omega=e^{i2\pi/3}$.
}
\label{tt22}
\end{table}
The relevant part of the  renormalizable superpotential involving
only the new chiral superfields is given by
\bea
 W \supset  Y_{nj} N^c_n (H_u L_j) + (\lambda_N)_{nm} S N_n N^0_m
     + m_{nm} N^c_n N_m+  \frac{1}{2} m^{\prime}_{nm} N^0_n N^0_m.
 \label{fig55}
\eea
For simplicity, we set $m_{nm} = m^{\prime}_{nm} =M_7\delta_{nm}$,
 $(\lambda_N)_{nm}= \lambda_N \delta_{ij}$ and  $Y_N\equiv Y_{nj}$.
Integrating out the new heavy chiral field and following the electroweak symmetry breaking,
 the light neutrino Majorana mass matrix is generated:
\bea
  m_\nu = \frac{(Y_N^T Y_N) v_u^2}{M_7} \times
          \frac{\lambda_N^T \lambda_N  \langle S\rangle^2}{M_7^2}.
\label{n77} \eea
We can see from this formula that the upper bound for seesaw scale
 is $M_7 \sim 10^6$ GeV, assuming all Yukawa coupling
 in Eq.~(\ref{fig55}) are ${\cal O}(1)$. It is clear from Eq.~(\ref{n77})
that we can have ${\cal O}$(1) $Y_N$ couplings  and the seesaw
scale $M_7$ around TeV  for $\lambda_N  \sim 10^{-4}$ or so. In this case the value for $\lambda_N$ is more natural
compared to the  dimension six case. Comparing Eq. (\ref{n77}) to  Eq. (\ref{eq6}), we can see that we have identical contributions to the lightest CP-even Higgs mass for effective dimension six and seven cases.

Having low ($\sim$TeV) scale for the inverse seesaw mechanism clearly makes the model accessible at the LHC. In Ref. \cite{Dev:2012zg} it is shown  that regions of the parameter space of the inverse seesaw model can be tested at the LHC, while Ref.~\cite{leptonflavor} shows that lepton flavor violation imposes strict constraints on these models.

\subsection{NMSSM+Triplets \label{nmssm+triplet}}

As pointed out in Ref. \cite{Gogoladze:2008wz},
another way for generating the dimension six operator is to
introduce  $SU(2)_L$  triplets with zero $(\Delta^c_0+\Delta_0)$ or with  unit $({\Delta^c_n}+{\Delta_n})$ hypercharge.
It was shown in Ref. \cite{Gogoladze:2008wz} that two pairs ($n=1,2$) of   $({\Delta^c_n}+{\Delta_n})$ are needed in order to generate the effective dimension  six operator for inverse seesaw mechanism. We will consider the case involving only $({\Delta^c_n}+{\Delta_n})$ as the additional fields. As an example we choose the charge assignments for the NMSSM fields shown as  $case \, I$ in Table \ref{table1}. Accordingly, in order to generate effective dimension six operators (see Eq. (\ref{sw2})) for the light neutrinos,  the $Z_3$ charges for $({\Delta^c_n}+{\Delta_n})$ fields are fixed, as given in Table \ref{tt3}.
The additional contributions to the NMSSM superpotential in this
case contain the following terms
\bea
 W \supset  Y_{ij} (L_i \Delta_1 L_j) + Y_{H_u} (H_u \Delta_2 H_u)
   + \lambda_N S \;
   {\rm tr}\left[ \bar{\Delta}_1 \bar{\Delta_2} \right]
   + m_1 \; {\rm tr}\left[ \bar{\Delta}_1  \Delta_1 \right]
   + m_2 \; {\rm tr}\left[ \bar{\Delta}_2    \Delta_2   \right],
   \label{hh2}
\eea
where $Y_{ij}$, $Y_{H_u}$ and $ \lambda_N$ are dimensionless Yukawa couplings and $m_1$, $m_2$ are mass parameters.
It is interesting to note that the interactions $(H_d \Delta_n H_d)$ and $(H_d \overline{\Delta}_n H_d)$, which can give significant contributions to the CP-even Higgs mass \cite{Espinosa:1998re}, are forbidden by $Z_3$ or $U(1)_Y$ symmetry.
\begin{table}[h]
\begin{center}
\renewcommand{\arraystretch}{1.2}
\begin{tabular}{| c|  c|  c | c | c | c | c | c | c| }
\hline
   & $SU(3)_c$ & $SU(2)_L$   & $U(1)_Y$ & $Z_3$ & $Z_2$ \\
\hline
 $\Delta_1$      &1  & 3  &   1  &  $1$  & $+$ \\
 \hline
 $\overline{\Delta}_1$         & 1 & 3  &   $-1$ &  $1$    & $+$ \\
  \hline
 $\Delta_2$      &1  & 3  &   $-1$  &  $\omega$  & $+$ \\
 \hline
 $\overline{\Delta}_2$   & 1 & 3  &   1 &  $\omega^2 $    & $+$ \\
 \hline
\end{tabular}
\end{center}
\caption{Charge assignments of $({\Delta_n}+{\overline{\Delta}_n})$ superfields, where $n=1,2$.   $\omega=e^{i2\pi/3}$ and   $Z_2$ is  matter parity.
}
\label{tt3}
\end{table}
%

\begin{figure}[b!]
\begin{center}
{\includegraphics*[width=.65\linewidth]{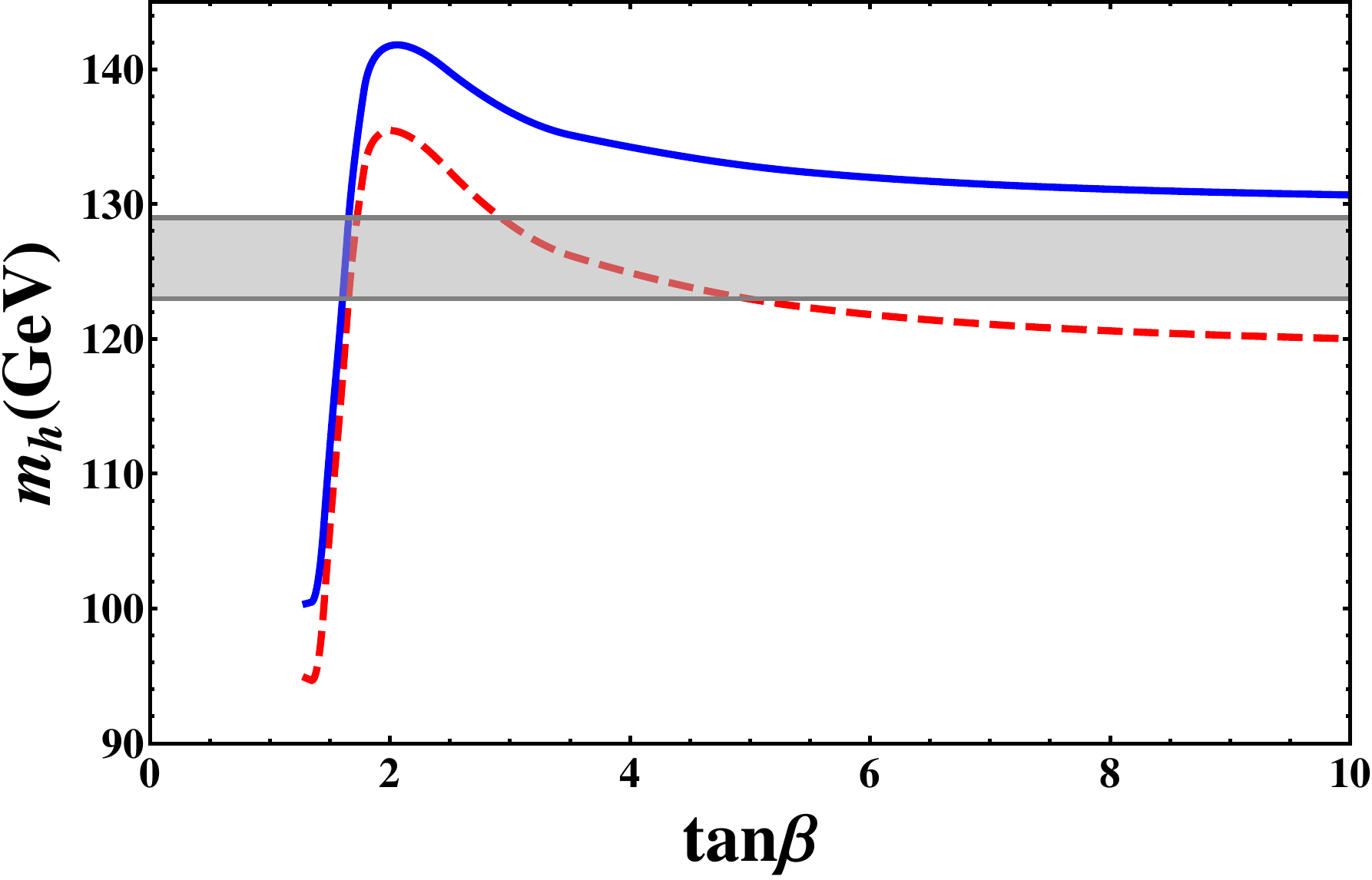}}
\caption{
Upper bounds on the lightest CP-even Higgs boson mass
versus $\tan \beta$, for  $M_S=200$ GeV, $\widetilde{X}_{t}=6$, $Y_{H_u}=0.15$, $m_1=m_2=3$ TeV. Maximum value of $\lambda$ is used. Red dashed line corresponds to NMSSM, and the blue solid line corresponds to NMSSM + $({\Delta_n}+{\overline{\Delta}_n})$. For reference, the gray band shows the Higgs mass range of $126\pm 3$ GeV.
}
\label{fig4}
\end{center}
\end{figure}

\begin{figure}[h]
\begin{center}
{\includegraphics*[width=.65\linewidth]{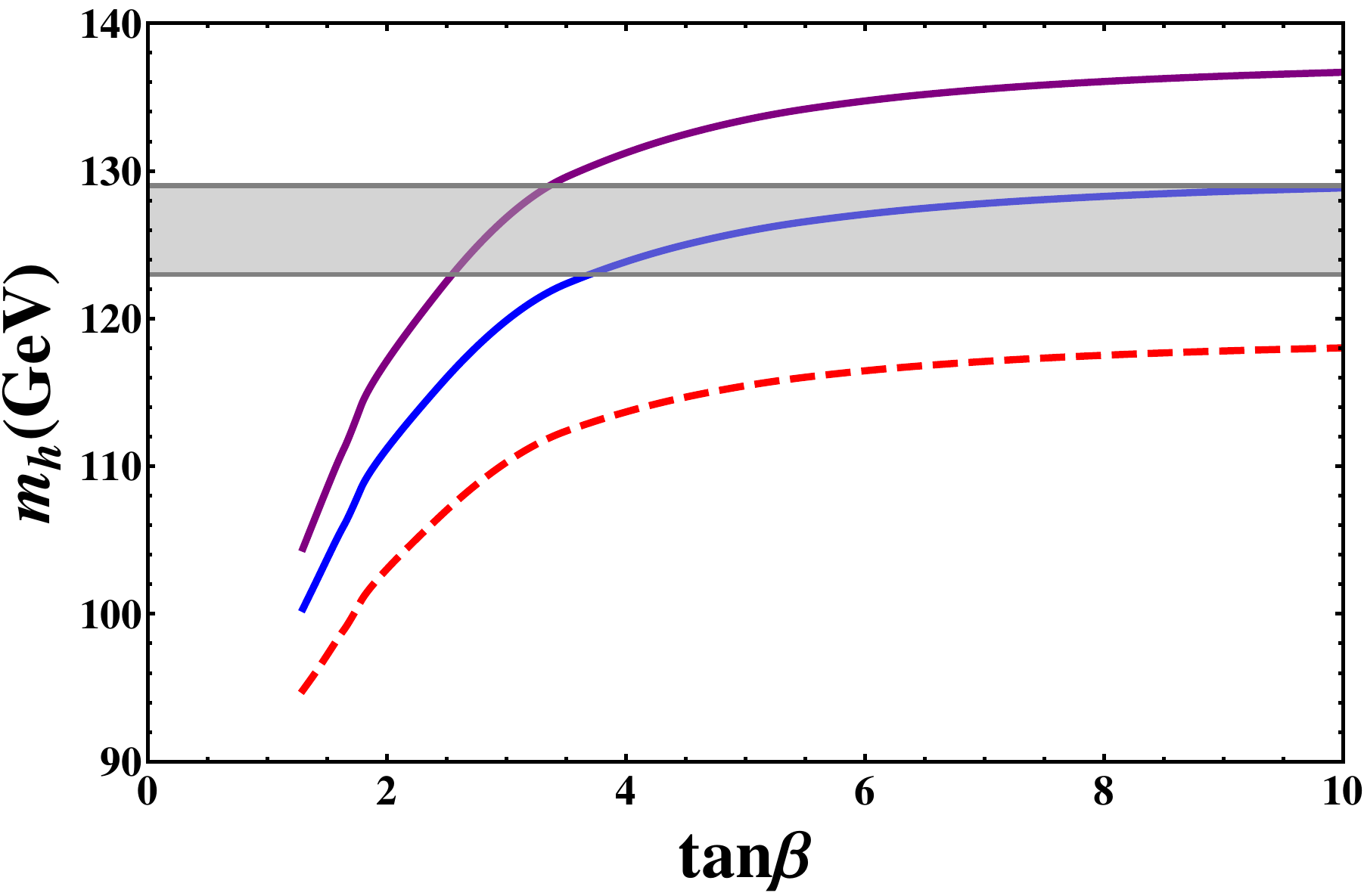}}
\caption{
Upper bounds on the lightest CP-even Higgs boson mass
versus $\tan \beta$, with  $M_S=200$ GeV, $\widetilde{X}_{t}=6$,  $m_1=m_2=3$ TeV, and  $\lambda = 0.3$. Red dashed line corresponds to the NMSSM, and blue and purple solid lines correspond to NMSSM + $({\Delta_n}+{\overline{\Delta}_n})$, with $Y_{H_u}=0.15$ and 0.2. For reference, the gray band shows the Higgs mass range of $126\pm 3$ GeV.
}
\label{fig5}
\end{center}
\end{figure}
%
The coupling $Y_H (H_u
\Delta H_u)$ in  Eq.~(\ref{hh2}) will generate a tree level
contribution to the lightest CP-even Higgs boson mass
 given by \cite{Espinosa:1998re}
\begin{eqnarray}
\left[ m_{h}^{2}\right] _{\Delta}=4 Y_{H_u}^2 v^2 \sin^4 \beta.
\label{e4}
\end{eqnarray}

We assume $\tan\beta\lesssim 50$ since for larger $\tan\beta$, there will be additional contribution in Eq. (\ref{e4}) which  can reduce the Higgs mass \cite{Ellwanger:2009dp}. To show the impact of $({\Delta_n}+{\overline{\Delta}_n})$ on the lightest CP-even Higgs mass, we plot in Figure \ref{fig4} the upper bounds on the Higgs mass versus $\tan\beta$.
We choose    $M_S=200$ GeV, $\widetilde{X}_{t}=6$, $Y_{H_u}=0.15$ and $m_1=m_2=3$ TeV. The red dashed line corresponds to NMSSM. The blue solid line corresponds to NMSSM+ $({\Delta_n}+{\overline{\Delta}_n})$. For reference, the gray band corresponds to a Higgs mass of $126\pm 3$ GeV. We see that there is no  need for very large, ${\cal O}(1)$, value for the coupling $Y_{H_u}$ in order to have a 126 GeV Higgs. As seen from Figure \ref{fig4}, $Y_{H_u}=0.15$ already yields an upper bound on the Higgs mass above 126 GeV. The upper bound is 142 GeV for $\tan\beta\approx 2$, which asymptotically approaches  $m_h \approx 130$ GeV for larger $\tan\beta$ values.
We are able to realize a Higgs mass of around 126 GeV with relatively small values of $\lambda$ and $Y_{H_u}$. We therefore conclude that in the NMSSM with the inverse seesaw mechanism for neutrinos, we can have relatively light, ${\cal O}$(200) GeV or so, stop quarks  without invoking maximal values for $\lambda$ or $Y_{H_u}$, and also without severely restricting $\tan\beta$.

Consider a case with relatively small value of $\lambda$. In Figure \ref{fig5} we choose $\lambda=$0.3 and $M_S=200$ GeV. The red dashed line corresponds to the Higgs mass bound in NMSSM, while the blue and purple lines show the bounds with $Y_{H_u}=0.15$ and $Y_{H_u}=0.2$. One can see that it is fairly easy to increase the Higgs mass bound in NMSSM to 126 GeV, even with small values of $\lambda$ and  $Y_{H_u}$.

Notice that we can use a triplet with zero hypercharge to generate the inverse seesaw operator. In this case the superpotential looks exactly the same,  with ($\Delta^c_0+\Delta_0$ ) replacing ($N_n^c+N_n$). The result will be similar to what we presented in section  \ref{nmssm+singlet}  when we consider 3 pairs of ($N_n^c+N_n$). Given this similarity we do not extend our analysis to the case of an $SU(2)_L$ triplet with zero hypercharge.

The low scale triplet model has a very nice feature. A light triplet  not only helps generate the inverse seesaw mechanism and provides significant contribution to the CP-even Higgs boson mass, it also contributes to the enhancement of the Higgs production and decay in the diphoton channel. In order to have a sizable effect on the diphoton production, the coupling involving the triplet and the Higgs doublets has to be large \cite{Delgado:2012sm}, which makes the coupling non-perturbative below the Planck or GUT scale. On the other hand, as was shown in Ref. \cite{Gogoladze:2008wz}, we need one pair of $SU(2)_L$ triplets  with zero hypercharge or two pairs of triplets with unit hypercharge. In both cases at least  one of the gauge couplings becomes  non-perturbative below  the GUT scale. However, if the theory is still valid near the Landau pole all couplings will become large through the two and higher loop renormalization group equations (RGEs). The couplings can effectively merge  together and we can have non-perturbative unification \cite{Maiani:1977cg}.
 Another attractive feature of the light triplet extension of the NMSSM (MSSM) spectrum  is that it can help resolve the little hierarchy problem \cite{Agashe:2011ia} with a Higgs mass of around 126 GeV.

\section{Conclusion \label{conclusions}}

Following Ref. \cite{Gogoladze:2008wz}, we consider extensions of the next-to-minimal supersymmetric model (NMSSM) in which the observed neutrino masses are
generated through TeV scale inverse seesaw mechanism. We have shown that the new particles associated with the inverse seesaw mechanism  can have sizable couplings to the lightest CP-even Higgs field which can yield a large contribution to its mass. This new contribution makes  it possible  to have a 126 GeV Higgs with order of 200 GeV stop quarks mass and a broad range of $\tan\beta$ values. This can be exploited to enhance the Higgs production and decay in the diphoton  channel as well.

\section*{Acknowledgments}
We thank   M. Adeel Ajaib, Aleksandr Azatov  and Mansoor Ur Rehman   for valuable discussions.
This work is supported in part by the DOE Grant No. DE-FG02-12ER41808. We also thank the Bartol Research Institute for partial support.



\begin{thebibliography}{99}

\bibitem{ATLAS:jul4}
F. Gianotti, CERN Seminar, ``Update on the Standard Model Higgs searches in ATLAS",
July, 4 2012. ATLAS-CONF-2012-093

\bibitem{CMS:jul4}
J. Incandela, CERN Seminar, ``Update on the Standard Model Higgs searches in CMS",
July, 4 2012. CMS-PAS-HIG-12-020.



\bibitem{mssm}
See  for instance  S.~P.~Martin,
  hep-ph/9709356 and references therein.

\bibitem{Ajaib:2012vc}
  M.~A.~Ajaib, I.~Gogoladze, F.~Nasir and Q.~Shafi,
  Phys.\ Lett.\ B {\bf 713}, 462 (2012).


\bibitem{Djouadi:2005gj}
For a review  see A.~Djouadi,
  Phys.\ Rept.\  {\bf 459}, 1 (2008) and reference therein.




  \bibitem{b5}
P.~H.~Chankowski, J.~R.~Ellis and S.~Pokorski,
Phys.\ Lett.\ B \textbf{423}, 327 (1998);
P.~H.~Chankowski, J.~R.~Ellis, M.~Olechowski and S.~Pokorski,
Nucl.\ Phys.\ B \textbf{544}, 39 (1999);
G.~L.~Kane and S.~F.~King, 
Phys.\ Lett.\ B \textbf{451}, 113 (1999);
G.~L.~Kane, J.~D.~Lykken, B.~D.~Nelson and L.~T.~Wang,
Phys.\ Lett.\ B \textbf{551}, 146 (2003).


\bibitem{Ellwanger:2009dp}
For a review  see  U.~Ellwanger, C.~Hugonie and A.~M.~Teixeira,
  Phys.\ Rept.\  {\bf 496}, 1 (2010);
  M.~Maniatis,
  Int.\ J.\ Mod.\ Phys.\ A {\bf 25}, 3505 (2010) and reference therein.



\bibitem{natural}
  U.~Ellwanger,
  JHEP {\bf 1203}, 044 (2012):
  L.~J.~Hall, D.~Pinner and J.~T.~Ruderman, JHEP {\bf 04}, 131 (2012);
  Z.~Kang, J.~Li and T.~Li,
  arXiv:1201.5305 [hep-ph];
  U.~Ellwanger, G.~Espitalier-Noel and C.~Hugonie,
  JHEP {\bf 1109}, 105 (2011);
  T.~Cheng, J.~Li, T.~Li, X.~Wan, Y.~k.~Wang and S.~-h.~Zhu,
  arXiv:1207.6392 [hep-ph];
  K.~Schmidt-Hoberg and F.~Staub,
  arXiv:1208.1683 [hep-ph];
  M.~Perelstein and B.~Shakya,
  arXiv:1208.0833 [hep-ph];
  J.~Cao, Z.~Heng, J.~M.~Yang, Y.~Zhang and J.~Zhu,
  arXiv:1202.5821 [hep-ph];
  J.~Cao, Z.~Heng, J.~M.~Yang and J.~Zhu,
  arXiv:1207.3698 [hep-ph].
  
  
 


\bibitem{Abe:2007kf}
  H.~Abe, T.~Kobayashi and Y.~Omura,
  Phys.\ Rev.\ D {\bf 76}, 015002 (2007);
  I.~Gogoladze, M.~U.~Rehman and Q.~Shafi,
  Phys.\ Rev.\ D {\bf 80}, 105002 (2009);
  D.~Horton and G.~G.~Ross,
  Nucl.\ Phys.\ B {\bf 830}, 221 (2010);
  J.~E.~Younkin and S.~P.~Martin,
  Phys.\ Rev.\ D {\bf 85}, 055028 (2012);
  S.~Antusch, L.~Calibbi, V.~Maurer, M.~Monaco and M.~Spinrath,
  arXiv:1207.7236 [hep-ph].



\bibitem{Gunion:2012zd}
  J.~F.~Gunion, Y.~Jiang and S.~Kraml,
  Phys.\ Lett.\ B {\bf 710}, 454 (2012);
  S.~F.~King, M.~Muhlleitner and R.~Nevzorov,
  Nucl.\ Phys.\ B {\bf 860}, 207 (2012);
  D.~A.~Vasquez, G.~Belanger, C.~Boehm, J.~Da Silva, P.~Richardson and C.~Wymant,
  arXiv:1203.3446 [hep-ph];
  U.~Ellwanger and C.~Hugonie,
  arXiv:1203.5048 [hep-ph];
  E.~Gabrielli, K.~Kannike, B.~Mele, A.~Racioppi and M.~Raidal,
  arXiv:1204.0080 [hep-ph];
  J.~Rathsman and T.~Rossler,
  arXiv:1206.1470 [hep-ph].



\bibitem{nonperturb}
 E.~Hardy, J.~March-Russell and J.~Unwin,
 arXiv:1207.1435 [hep-ph].


\bibitem{Gogoladze:2008wz}
  I.~Gogoladze, N.~Okada and Q.~Shafi,
  Phys.\ Lett.\ B {\bf 672}, 235 (2009).



\bibitem{InvSeesaw}
R.~N.~Mohapatra,
  Phys.\ Rev.\ Lett.\  {\bf 56} (1986) 561;
%
R.~N.~Mohapatra and J.~W.~F.~Valle,
  Phys.\ Rev.\  D {\bf 34}, 1642 (1986).



\bibitem{Djouadi:1998az}
  A.~Djouadi,
  Phys.\ Lett.\ B {\bf 435}, 101 (1998);
  R.~Dermisek and I.~Low,
  Phys.\ Rev.\ D {\bf 77}, 035012 (2008);
  I.~Low and S.~Shalgar,
  JHEP {\bf 0904}, 091 (2009).
  I.~Low, R.~Rattazzi and A.~Vichi,
  JHEP {\bf 1004}, 126 (2010);
 I.~Low and A.~Vichi,
  Phys.\ Rev.\ D {\bf 84}, 045019 (2011).



\bibitem{Desai:2012qy}
N.~Desai, B.~Mukhopadhyaya and S.~Niyogi,
  arXiv:1202.5190 [hep-ph];
 D.~Carmi, A.~Falkowski, E.~Kuflik and T.~Volansky,
  arXiv:1202.3144 [hep-ph];
  N.~D.~Christensen, T.~Han and S.~Su,
  arXiv:1203.3207 [hep-ph];
  R.~Benbrik, M.~G.~Bock, S.~Heinemeyer, O.~Stal, G.~Weiglein and L.~Zeune,
  arXiv:1207.1096 [hep-ph];
  N.~Arkani-Hamed, K.~Blum, R.~T.~D'Agnolo and J.~Fan,
  arXiv:1207.4482 [hep-ph];
  A.~Joglekar, P.~Schwaller and C.~E.~M.~Wagner,
  arXiv:1207.4235 [hep-ph];
 M.~R.~Buckley and D.~Hooper,
  arXiv:1207.1445 [hep-ph];
 Z.~Kang, T.~Li, J.~Li and Y.~Liu,
  arXiv:1208.2673 [hep-ph].


\bibitem{Ajaib:2012eb}
  M.~A.~Ajaib, I.~Gogoladze and Q.~Shafi,
  arXiv:1207.7068 [hep-ph].

\bibitem{Panagiotakopoulos:1998yw}
  C.~Panagiotakopoulos and K.~Tamvakis,
  Phys.\ Lett.\ B {\bf 446}, 224 (1999);
  C.~Panagiotakopoulos and A.~Pilaftsis,
  Phys.\ Rev.\ D {\bf 63}, 055003 (2001).


\bibitem{nmssm}
  G.~K.~Yeghian,
  hep-ph/9904488;
  U.~Ellwanger and C.~Hugonie,
  Eur.\ Phys.\ J.\ C {\bf 25}, 297 (2002).




\bibitem{Degrassi:2002fi}
  G.~Degrassi, S.~Heinemeyer, W.~Hollik, P.~Slavich and G.~Weiglein,
  Eur.\ Phys.\ J.\ C {\bf 28}, 133 (2003).

\bibitem{tjli}
  Z.~Kang and T.~Li, 
  arXiv:1111.7313 [hep-ph]


\bibitem{Babu:2004xg}
  K.~S.~Babu, I.~Gogoladze and C.~Kolda,
  hep-ph/0410085;
  K.~S.~Babu, I.~Gogoladze, M.~U.~Rehman and Q.~Shafi,
  Phys.\ Rev.\ D {\bf 78}, 055017 (2008).



\bibitem{Dev:2012zg}
  P.~S.~B.~Dev, R.~Franceschini and R.~N.~Mohapatra,
  arXiv:1207.2756 [hep-ph].
  A.~Das and N.~Okada,
  arXiv:1207.3734 [hep-ph];
  P.~Bandyopadhyay, E.~J.~Chun, H.~Okada and J.~-C.~Park,
  arXiv:1209.4803 [hep-ph].

\bibitem{leptonflavor}
    M.~Hirsch, W.~Porod, L.~Reichert and F.~Staub,
    arXiv:1206.3516 [hep-ph];
    A.~Abada, D.~Das, A.~Vicente, and C.~Weiland,
    arXiv:1206.6497 [hep-ph].
    
   



\bibitem{Espinosa:1998re}
  J.~R.~Espinosa and M.~Quiros,
  Phys.\ Rev.\ Lett.\  {\bf 81}, 516 (1998).


\bibitem{Delgado:2012sm}
  A.~Delgado, G.~Nardini and M.~Quiros,
  arXiv:1207.6596 [hep-ph].


\bibitem{Maiani:1977cg}
  L.~Maiani, G.~Parisi and R.~Petronzio,
  Nucl.\ Phys.\ B {\bf 136} (1978) 115;
    N.~Cabibbo and G.~R.~Farrar,
  Phys.\ Lett.\ B {\bf 110} (1982) 107.



\bibitem{Agashe:2011ia}
  K.~Agashe, A.~Azatov, A.~Katz and D.~Kim,
  Phys.\ Rev.\ D {\bf 84}, 115024 (2011);
  T.~Basak and S.~Mohanty,
  arXiv:1204.6592 [hep-ph];
  K.~Agashe, Y.~Cui and R.~Franceschini,
  arXiv:1209.2115 [hep-ph].









\end{thebibliography}
\end{document}